\documentclass[11pt,english,onecolumn]{IEEEtran}
\usepackage[T1]{fontenc}
\usepackage[latin9]{inputenc}
\usepackage[letterpaper]{geometry}
\usepackage{color}
\usepackage{float}
\usepackage{amsmath}
\usepackage{graphicx}
\usepackage{amssymb}

\makeatletter

\newcommand{\lyxline}[1][1pt]{%
  \par\noindent%
  \rule[.5ex]{\linewidth}{#1}\par}
\floatstyle{ruled}
\newfloat{algorithm}{tbp}{loa}
\floatname{algorithm}{Algorithm}

\newtheorem{thm}{Theorem}
\newtheorem{lemma}{Lemma}

\usepackage{tikz}
\usepackage{psfrag}

\makeatother

\usepackage{babel}

\begin{document}

\title{{\Large Message-Passing Inference on a Factor Graph for Collaborative
Filtering}}

\author{Byung-Hak Kim, Arvind Yedla, and Henry D. Pfister\\
{\normalsize Department of Electrical and Computer Engineering,
Texas A\&M University}\\
{\normalsize College Station, TX 77843, USA }\\
{\normalsize \{bhkim, yarvind, hpfister\}@tamu.edu}}
\maketitle
\begin{abstract}
This paper introduces a novel message-passing (MP) framework for the
collaborative filtering (CF) problem associated with recommender systems.
We model the movie-rating prediction problem popularized by the Netflix
Prize, using a probabilistic factor graph model and study the model
by deriving generalization error bounds in terms of the training error.
Based on the model, we develop a new MP algorithm, termed IMP, for
learning the model. To show superiority of the IMP algorithm, we compare
it with the closely related expectation-maximization (EM) based algorithm
and a number of other matrix completion algorithms. Our simulation
results on Netflix data show that, while the methods perform similarly
with large amounts of data, the IMP algorithm is superior for small
amounts of data. This improves the cold-start problem of the CF systems
in practice. Another advantage of the IMP algorithm is that it can
be analyzed using the technique of density evolution (DE) that was
originally developed for MP decoding of error-correcting codes. \\

\end{abstract}
\vspace{-4mm}

\section{Introduction\label{sec:intro}}

One compelling application of collaborative filtering is the automatic
generation of recommendations. For example, the Netflix Prize \cite{netflix}
has increased the interest in this field dramatically. Recommendation
systems analyze, in essence, patterns of user interest in items to
provide personalized recommendations of items that might suit a user's
taste. Their ability to characterize and recommend items within huge
collections has been steadily increasing and now represents a computerized
alternative to human recommendations. In the collaborative filtering,
the recommender system would identify users who share the same preferences
(e.g. rating patterns) with the active user, and propose items which
the like-minded users favored (and the active user has not yet seen).
One difficult part of building a recommendation system is accurately
predicting the preference of a user, for a given item, based only
on a few known ratings. The collaborative filtering problem is now
being studied by a broad research community including groups interested
in statistics, machine learning and information theory \cite{KDD07,Lafferty stat}.
Recent works on the collaborative filtering problem can be largely
divided into two areas:
\begin{enumerate}
\item The first area considers efficient models and practical algorithms.
There are two primary approaches: \emph{neighborhood} model approaches
that are loosely based on {}``$k$-Nearest Neighbor'' algorithms
and \emph{factor} models (e.g., low dimension or low-rank models with
a least squares flavor) such as hard clustering based on singular
vector decomposition (SVD) or probabilistic matrix factorization (PMF)
and soft clustering which employs expectation maximization (EM) frameworks
\cite{KDD07,PMF1,PMF2,Hofmann uai99,hak tech}. 
\item The second area involves exploration of the fundamental limits of
these systems. Prior work has developed some precise relationships
between sparse observation models and the recovery of missing entries
in terms of the matrix completion problem under the restriction of
low-rank matrices model or clustering models \cite{LowRank,Onkar,SparseObs}.
This area is closely related with the practical issues known as cold-start
problem \cite{Cold,netflix}. That is, giving recommendations to new
users who have submitted only a few ratings, or recommending new items
that have received only a few ratings from users. In other words,
how few ratings to be provided for the the system to guess the preferences
and generate recommendations?
\end{enumerate}
In this paper, we employ an alternative modern coding-theoretic approach
that have been very successful in the field of error-correcting codes
to the problem. Our results are different from the above works in
several aspects as outlined below.
\begin{enumerate}
\item Our approach tries to combine the benefits of clustering users and
movies into groups probabilistically and applying a factor analysis
to make predictions based on the groups. The precise probabilistic
generative factor graph model is stated and generalization error bounds
of the model with some observations are studied in Sec. \ref{sec:factor}.
Based on the model, we derive a MP based algorithms, termed IMP, which
has demonstrated empirical success in other applications: low-density
parity-check codes and turbo-codes decoding. Furthermore, as a benchmark,
popular EM algorithms which are frequently used in both learning and
coding community \cite{Hofmann uai99,Mackay,Neal Graph} are developed
in Sec. \ref{sec:algs}.
\item Our goal is to characterize system limits via modern coding-theoretic
techniques. Toward this end, we provide a characterization of the
messages distribution passed on the graph via density evolution (DE)
in Sec. \ref{sec:DE_sec}. DE is an asymptotic analysis technique
that was originally developed for MP decoding of error-correcting
codes. Also, through the emphasis of simulations on cold-start settings,
we see the cold start problem is greatly reduced by the IMP algorithm
in comparison to other methods on real Netflix data com in Sec. \ref{sec:Simulations}.
\end{enumerate}

\section{Factor graph model\label{sec:factor}}

\subsection{Model Description}

Consider a collection of $N$ users and $M$ movies when the set $O$
of user-movie pairs have been observed. The main theoretical question
is, {}``How large should the size of $O$ be to estimate the unknown
ratings within some distortion $\delta$?''. Answers to this question
certainly require some assumptions about the movie rating process
as been studied by prior works \cite{LowRank,Onkar}. So we begin
differently by introducing a probabilistic model for the movie ratings.
The basic idea is that \emph{hidden} variables are introduced for
users and movies, and that the movie ratings are conditionally independent
given these hidden variables. It is convenient to think of the hidden
variable for any user (or movie) as the \emph{user group} (or \emph{movie
group}) of that user (or movie). In this context, the rating associated
with a user-movie pair depends only on the user group and the movie
group. 

Let there be $g_{u}$ user groups, $g_{v}$ movie groups, and define
$[k]\triangleq\left\{ 1,2,\ldots,k\right\} $. The user group of the
$n$-th user, $U_{n}\in[g_{u}]$, is a discrete r.v. drawn from $\Pr(U_{n}=u)\triangleq p_{U}(u)$
and $\mathbf{U}=U_{1},U_{2},\ldots,U_{N}$ is the user group vector.
Likewise, the movie group of the $m$-th movie, $V_{m}\in[g_{v}]$,
is a discrete r.v. drawn from $\Pr(V_{m}=v)\triangleq p_{V}(v)$ and
$\mathbf{V}=V_{1},V_{2},\ldots,V_{M}$ is the movie group vector.
Then, the rating of the $m$-th movie by the $n$-th user is a discrete
r.v. $R_{nm}\in\mathcal{R}$ (e.g., Netflix uses $\mathcal{R}=[5]$)
drawn from $\Pr(R_{nm}=r|U_{n}=u,V_{m}=v)\triangleq w(r|u,v)$ and
the rating $R_{nm}$ is \emph{conditionally independent} given the
user group $U_{n}$ and the movie group $V_{m}$. Let $\mathbf{R}$
denote the rating matrix and the observed submatrix be $\mathbf{R}_{O}$
with $O\subseteq[N]\times[M]$. In this setup, some of the entries
in the rating matrix are observed while others must be predicted.
The conditional independence assumption in the model implies that
\[
\Pr\left(\mathbf{R}_{O}|\mathbf{U},\mathbf{V}\right)\triangleq\prod_{(n,m)\in O}w\left(R_{nm}|U_{n},V_{m}\right).\]

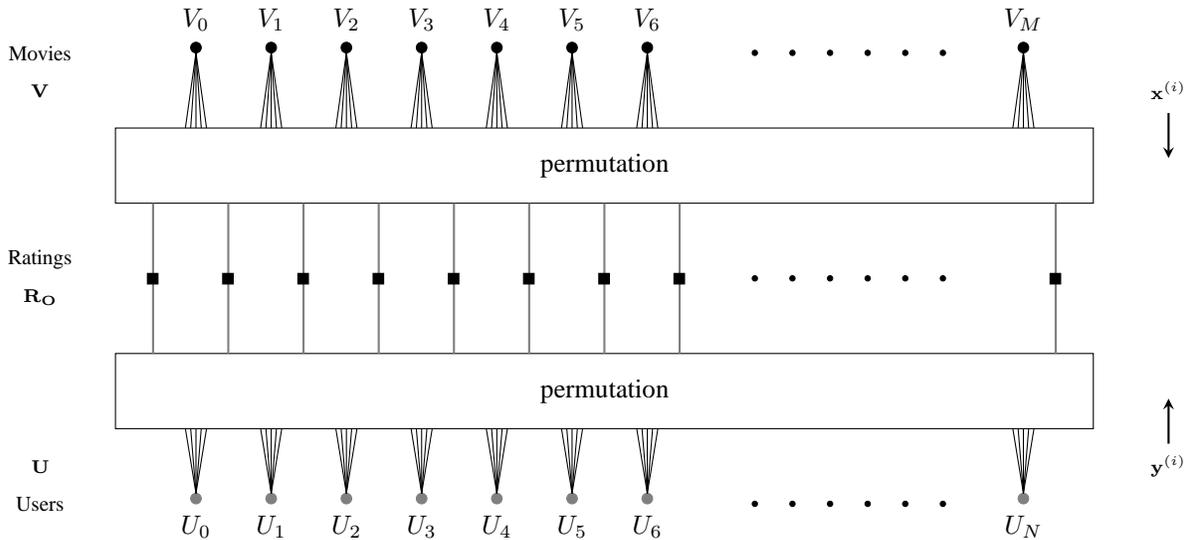
\begin{figure*}[t]
{\small \centering 
\begin{tikzpicture}[>=stealth,scale=1.00]
\draw (-1,2) rectangle +(13,1);
\draw (5.5,2.5) node {permutation};
\draw (-1,5) rectangle +(13,1);
\draw (5.5,5.5) node {permutation};
\foreach \x/\userlabel/\movielabel in {0/0/0,1/1/1,2/2/2,3/3/3,4/4/4,5/5/5,6/6/6,11/N/M}
{
\filldraw[gray] (\x,1)+(2pt,2pt) circle (2pt) node[black,below=3pt] {$U_{\userlabel}$};
\draw (\x,1)+(2pt,4pt) -- (\x,2);
\draw (\x,1)+(2pt,4pt) -- ([xshift = 4pt]\x,2);
\draw (\x,1)+(2pt,4pt) -- ([xshift = 2pt]\x,2);
\draw (\x,1)+(2pt,4pt) -- ([xshift = -2pt]\x,2);
\draw (\x,1)+(2pt,4pt) -- ([xshift = 6pt]\x,2);

\draw[thick,gray] (\x,3)+(0.5cm,0pt) -- ([xshift=0.5cm]\x,5);

\filldraw (\x,7)+(2pt,2pt) circle (2pt) node[black,above=3pt] {$V_{\movielabel}$};
\draw (\x,7)+(2pt,0pt) -- (\x,6);
\draw (\x,7)+(2pt,0pt) -- ([xshift = 4pt]\x,6);
\draw (\x,7)+(2pt,0pt) -- ([xshift = 2pt]\x,6);
\draw (\x,7)+(2pt,0pt) -- ([xshift = -2pt]\x,6);
\draw (\x,7)+(2pt,0pt) -- ([xshift = 6pt]\x,6);
}

\draw[thick,gray] (-1,3)+(0.5cm,0pt) -- ([xshift=0.5cm]-1,5);

\foreach \x in {0.425,1.425,2.425,...,6.425,11.425} {
	\filldraw (\x,3.925) rectangle +(4pt,4pt);
}
\filldraw (-0.575,3.925) rectangle +(4pt,4pt);
\foreach \x in {7.5,8,8.5,...,10} {
\foreach \y in {1,4,7} {
\filldraw (\x,\y) circle (1pt);
}}

\draw (-2,1.5) node {$\scriptstyle\mathbf{U}$};
\draw (13,1.5) node {$\scriptstyle\mathbf{y}^{(i)}$};
\draw[->,thick] (13,1.8) -- (13,2.4);
\draw (-2,1) node {\scriptsize{Users}};
\draw (-2,4.25) node {\scriptsize{Ratings}};
\draw (-2,3.75) node {$\scriptstyle\mathbf{R_{O}}$};
\draw (-2,6.5) node {$\scriptstyle\mathbf{V}$};
\draw (13,6.5) node {$\scriptstyle\mathbf{x}^{(i)}$};
\draw[->,thick] (13,6.2) -- (13,5.6);
\draw (-2,7) node {\scriptsize{Movies}};
\end{tikzpicture}

}{\small \par}

{\small \caption{\label{fig:tanner}The factor graph model for the collaborative filtering
problem. The graph is sparse when there are few ratings. Edges represent
random variables and nodes represent local probabilities. The node
probability associated with the ratings implies that each rating depends
only on the movie group (top edge) and the user group (bottom edge).
Synthetic data can be generated by picking i.i.d. random user/movie
groups and then using random permutations to associate groups with
ratings. Note $\mathbf{x}^{(i)}$ and $\mathbf{y}^{(i)}$ are the
messages from movie to user and user to movie during iteration $i$
for the Algorithm \ref{alg:IMP}.}
}
\end{figure*}

Specifically, we consider the factor graph (composed of 3 layers,
see Fig. \ref{fig:tanner}) as a randomly chosen instance of this
problem based on this probabilistic model. The key assumptions are
that these layers separate the influence of user groups, movie groups,
and observed ratings and the outgoing edges from each user node are
attached to movie nodes via a random permutation. 

{\small }%
{\small \par}

The main advantage of our model is that, since it exploits the correlation
in ratings based on similarity between users (and movies) and includes
noise process, this model approximates real Netflix data generation
process more closely than other simpler factor models. It is also
important to note that this is a \emph{probabilistic generative model}
which allows one to evaluate different learning algorithms on synthetic
data and compare the results with theoretical bounds (see Sec. \ref{subsec:Discusssions}
for details).

\subsection{Generalization Error Bound }

\noindent In this section, we consider bounds on generalization from
partial knowledge of the (binary-rating) matrix for collaborative
filtering application. The tighter bound implies one can use most
of known ratings for learning the model completely. Since computation
of $\mathbf{R}$ can be viewed as the product of three matrices, we
consider the simplified class of tri-factorized matrices $\chi_{g_{u},g_{v}}$
as, \[
\left\{ X|X\,=\, U^{T}WV,U\,\,\in\,\,\left[0,\,1\right]^{g_{u}\times N}\,\,\,,V\,\,\in\,\,\left[0,\,1\right]^{g_{v}\times M}\,\,\,,W\,\,\in\,\,\left\{ \pm1\right\} ^{g_{u}\times g_{v}}\,\right\} .\]
We bound the overall distortion between the entire predicted matrix
$X$ and the true matrix $Y$ as a function of the distortion on the
observed set of size $|O|$ and the error $\epsilon$. Let $y\in\left\{ \pm1\right\} $
be binary ratings and define a zero-one sign agreement distortion
as

\[
d\left(x,\, y\right)\triangleq\begin{cases}
1 & \mbox{if}\, xy\leq0\\
0 & \mbox{otherwise}\end{cases}.\]
Also, define the average distortion over the entire prediction matrix
as \[
D\left(X,\, Y\right)\triangleq\sum_{(n,m)\in[N]\times[M]}d\left(x,\, y\right)/NM\]
 and the averaged observed distortion as \[
D_{O}\left(X,\, Y\right)\triangleq\sum_{(n,m)\in O}d\left(x,\, y\right)/|O|.\]
 
\begin{thm}
\label{thm:bound}For any matrix $Y\in\left\{ \pm1\right\} ^{N\times M}$,
$N,\, M>2$, $\delta>0$ and integers $g_{u}$ and $g_{v}$, with
probability at least $1-\delta$ over choosing a subset $O$ of entries
in $Y$ uniformly among all subsets of $|O|$ entries $\forall X\in\chi_{g_{u},g_{v}}$,
$|D\left(X,\, Y\right)-D_{O}\left(X,\, Y\right)|$ is upper bounded
by\\

\end{thm}
\[
\sqrt{\left\{ (Ng_{u}+Mg_{v}+g_{u}g_{v})\,\mbox{log}\,\frac{12eM}{\mbox{min}(g_{u},\, g_{v})}-\mbox{log}\delta\right\} /2|O|}\triangleq h\left(g_{u},\, g_{v},\, N,\, M,\,|O|\right).\]

\begin{proof}
The proof of this theorem is given in Appendix \ref{sec:Bound}. 
\end{proof}
Let us finish this section with two implications of the Thm. \ref{thm:bound}
in terms of the five parameters: $g_{u},\, g_{v},\, N,\, M,\,|O|$.
\begin{enumerate}
\item For fixed group numbers $g_{u}$ and $g_{v}$, as number of users
$N$ and movies $M$ increases, to keep the bound tight, number of
observed ratings $|O|$ also needs to grow in the same order.
\item For a fixed sized matrix, when the choice of $g_{u}$ and/or $g_{v}$
increases, $|O|$ needs to grow in the same order to prevent over-learning
the model. Also, as $|O|$ increases, we could increase the value
of $g_{u}$ and/or $g_{v}$. 
\end{enumerate}

\section{Learning Algorithms \label{sec:algs}}

\subsection{Message Passing (MP) Learning }

\noindent Once a generative model describing the data has been specified,
we describe how two algorithms can be applied in the model using a
unified cost function, the free energy. Since exact learning and inference
are often intractable, so we turn to approximate algorithms that search
distributions that are close to the correct posterior distribution
by minimizing pseudo-distances on distributions, called free energies
by statistical physicists. The problem can be formulated via message-passing
(also known as belief propagation) framework via the sum-product algorithm
since fixed points of (loopy) belief propagation correspond to extrema
of the Bethe approximation of the free energy \cite{Yedidia}. The
basic idea is that the local neighborhood of any node in the factor
graph is tree-like, so that belief propagation gives a nearly optimal
estimate of the a posteriori distributions. We denote the message
from movie $m$ to user $n$ during iteration $i$ by $\mathbf{x}_{m\to n}^{(i)}$
and the message from user $n$ to movie $m$ by $\mathbf{y}_{n\to m}^{(i)}$.
The set of all users whose rating movie $m$ was observed is denoted
$\mathcal{U}_{m}$ and the set of all movies whose rating by user
$n$ was observed is denoted $\mathcal{V}_{n}$. %
\begin{algorithm*}[t]
\caption{\label{alg:IMP}IMP Algorithm}

\textbf{Step I:} Initialization \[
\mathbf{x}_{m\to n}^{(0)}(v)\!=\!\mathbf{x}_{m}^{(0)}(v)\!=\! p_{V}(v),\,\mathbf{y}_{n\to m}^{(0)}(u)\!=\!\mathbf{y}_{n}^{(0)}(u)\!=\! p_{U}(u),\, w\left(r|u,v\right)\]

\textbf{Step II:} Recursive update 

\[
\mathbf{y}_{n\to m}^{(i+1)}(u)\!=\!\frac{\mathbf{y}_{n}^{(0)}(u)\!{\displaystyle \prod_{k\in\mathcal{V}_{n}\backslash m}}\!{\displaystyle \sum_{v}}w\left(r_{n,m}|u,v\right)\mathbf{x}_{k\to n}^{(i)}(v)}{{\displaystyle \sum_{u'}}\mathbf{y}_{n}^{(0)}(u')\!{\displaystyle \prod_{k\in\mathcal{V}_{n}\backslash m}}\!{\displaystyle \sum_{v}}w\left(r_{n,m}|u',v\right)\mathbf{x}_{k\to n}^{(i)}(v)}\]
\[
\mathbf{x}_{m\to n}^{(i+1)}(v)\!=\!\frac{\mathbf{x}_{m}^{(0)}(v){\displaystyle \!\prod_{k\in\mathcal{U}_{m}\backslash n}}\!{\displaystyle \sum_{u}}w\left(r_{n,m}|u,v\right)\mathbf{y}_{k\to m}^{(i)}(u)}{{\displaystyle \sum_{v'}}\mathbf{x}_{m}^{(0)}(v')\!{\displaystyle \prod_{k\in\mathcal{U}_{m}\backslash n}}\!{\displaystyle \sum_{u}}w\left(r_{n,m}|u,v'\right)\mathbf{y}_{k\to m}^{(i)}(u)}\]

\textbf{Step III:} Output\[
{\textstyle \hat{p}_{R_{nm}|\mathbf{R}_{O}}^{(i+1)}(r)}\!=\!\frac{{\displaystyle \sum_{u,v}}\mathbf{y}_{n\to m}^{(i+1)}(u)\mathbf{x}_{m\to n}^{(i+1)}(v)w\left(r|u,v\right)}{{\displaystyle \sum_{r}}{\displaystyle \sum_{u,v}}\mathbf{y}_{n\to m}^{(i+1)}(u)\mathbf{x}_{m\to n}^{(i+1)}(v)w\left(r|u,v\right)}\]
\[
{\textstyle \hat{p}_{U_{n}|\mathbf{R}_{O}}^{(i+1)}(u)}\!=\!\frac{\mathbf{y}_{n}^{(0)}(u)\!{\displaystyle \prod_{k\in\mathcal{V}_{n}}}\!{\displaystyle \sum_{v}}w\left(r_{n,m}|u,v\right)\mathbf{x}_{k\to n}^{(i)}(v)}{{\displaystyle \sum_{u'}}\mathbf{y}_{n}^{(0)}(u')\!{\displaystyle \prod_{k\in\mathcal{V}_{n}}}\!{\displaystyle \sum_{v}}w\left(r_{n,m}|u',v\right)\mathbf{x}_{k\to n}^{(i)}(v)}\]
\[
\hat{p}_{V_{m}|\mathbf{R}_{O}}^{(i+1)}(v)\!=\!\frac{\mathbf{x}_{m}^{(0)}(v){\displaystyle \!\prod_{k\in\mathcal{U}_{m}}}\!{\displaystyle \sum_{u}}w\left(r_{n,m}|u,v\right)\mathbf{y}_{k\to m}^{(i)}(u)}{{\displaystyle \sum_{v'}}\mathbf{x}_{m}^{(0)}(v')\!{\displaystyle \prod_{k\in\mathcal{U}_{m}}}\!{\displaystyle \sum_{u}}w\left(r_{n,m}|u,v'\right)\mathbf{y}_{k\to m}^{(i)}(u)}\]
 
\end{algorithm*}
\begin{algorithm*}[t]
\caption{\label{alg:EM_alg}EM Learning Algorithm}

\textbf{Step I:} Initialization\[
f_{n}^{(0)}(u)=p_{U}(u),\, h_{m}^{(0)}(v)=p_{V}(v),\, w^{(0)}\left(r|u,v\right)\]

\textbf{Step II:} Recursive update\begin{align*}
f_{n}^{(i+1)}(u) & \!=\!\frac{{\displaystyle \sum_{m\in\mathcal{V}_{n}}}f_{n}^{(i)}\left(u\right)\!{\displaystyle \sum_{v\in[g_{m}]}}\! w^{(i)}\left(r_{n,m}|u,v\right)h_{m}^{(i)}(v)}{{\displaystyle \sum_{u'\in[g_{u}]}}{\displaystyle \sum_{m\in\mathcal{V}_{n}}}f_{n}^{(i)}\left(u\right)\!{\displaystyle \sum_{v\in[g_{m}]}}\! w^{(i)}\left(r_{n,m}|u,v\right)h_{m}^{(i)}(v)}\end{align*}
\[
h_{m}^{(i+1)}(v)\!=\!\frac{{\displaystyle \sum_{n\in\mathcal{U}_{m}}}h_{m}^{(i)}(v)\!{\displaystyle \sum_{u\in[g_{u}]}}\! w^{(i)}\left(r_{n,m}|u,v\right)f_{n}^{(i)}\left(u\right)}{{\displaystyle \sum_{v'\in[g_{v}]}}{\displaystyle \sum_{n\in\mathcal{U}_{m}}}h_{m}^{(i)}(v)\!{\displaystyle \sum_{u\in[g_{u}]}}\! w^{(i)}\left(r_{n,m}|u,v\right)f_{n}^{(i)}\left(u\right)}\]
\[
w^{(i+1)}\left(r|u,v\right)\!=\!\frac{{\displaystyle \sum_{(n,m):r_{n,m}=r}}w^{(i)}\left(r_{n,m}|u,v\right)f_{n}^{(i+1)}(u)h_{m}^{(i+1)}(v)}{{\displaystyle \sum_{r\in\mathcal{R}}}{\displaystyle \sum_{(n,m):r_{n,m}=r}}w^{(i)}\left(r_{n,m}|u,v\right)f_{n}^{(i+1)}(u)h_{m}^{(i+1)}(v)}\]

\textbf{Step III:} Output\[
{\textstyle \hat{p}_{R_{nm}|\mathbf{R}_{O}}^{(i+1)}(r)}\!=\!\frac{{\displaystyle \sum_{u,v}}\! f_{n}^{(i+1)}(u)h_{m}^{(i+1)}(v)w^{(i+1)}\left(r|u,v\right)}{{\displaystyle \sum_{r\in\mathcal{R}}}{\displaystyle \sum_{u,v}}\! f_{n}^{(i+1)}(u)h_{m}^{(i+1)}(v)w^{(i+1)}\left(r|u,v\right)}\]
\[
{\textstyle \hat{p}_{U_{n}|\mathbf{R}_{O}}^{(i+1)}(u)}\!=\! f_{n}^{(i+1)}(u)\]
 \[
\hat{p}_{V_{m}|\mathbf{R}_{O}}^{(i+1)}(v)\!=\! h_{m}^{(i+1)}(v)\]

\end{algorithm*}
\begin{algorithm*}[t]
\caption{\label{alg:CR_alg}{\small VDVQ Clustering Algorithm via GLA Splitting
(shown only for users)}}

\textbf{Step I:} Initialization\textcolor{black}{{} }

~~\textcolor{black}{Let $i=j=0$ }and $c_{m}^{(0,0)}(0)$\textcolor{black}{~be
the average rating of movie $m$.}\\

\textbf{Step II:} Splitting of critics

~~Set \[
c_{m}^{(i+1,j)}(u)\!=\!\begin{cases}
c_{m}^{(i,j)}(u) & u\!=\!0,\ldots,2^{i}\!-\!1\\
c_{m}^{(i,j)}(u\!-\!2^{i})\!+\! z_{m}^{(i+1,j)}(u) & u\!=\!2^{i},\ldots,2^{i+1}\!\!-\!1\end{cases}\]
where the $z_{m}^{(i+1,j)}(u)$ are i.i.d. random variables with small
variance.\\

\textbf{Step III:} Recursive soft K-means clustering for $c_{m}^{(i,j)}(u)$
for $j=1,\,\ldots\,,\, J$.

~~1. Each training data is assigned a soft degree of assignment
$\pi_{n}\left(u\right)$ to each of the critics using \[
\pi_{n}^{(i,j)}\left(u\right)=\frac{\mbox{exp}\left(-\beta d\left(\mathbf{R}_{O},c_{m}^{(i,j)}(u)\right)\right)}{{\displaystyle \sum_{u'\in[g_{u}]}}\mbox{exp}\left(-\beta d\left(\mathbf{R}_{O},c_{m}^{(i,j)}(u')\right)\right)}\]
where $d\left(\mathbf{R}_{O},c_{m}^{(i,j)}(u)\right)=\sqrt{\sum_{(n,m)\in O}\left(c_{nm}^{(i,j)}(u)-r_{n,m}\right)^{2}\!\!/\!|O|}$
, $g_{u}=2^{i+1}$.

~~2. Update all critics as

\[
c_{m}^{(i,j+1)}(u)=\frac{\sum_{n}\pi_{n}^{(i,j)}\left(u\right)c_{m}^{(i,j)}(u)}{\sum_{n}\pi_{n}^{(i,j)}\left(u\right)}.\]
\\
\textbf{Step IV:} Repeat Steps II and III until the desired number
of critics $g_{u}$ is obtained. 

\textbf{Step V:} Estimate of $w(r|u,v)$

~~After clustering users/movies each into user/movie groups with
the soft group membership $\pi_{n}\left(u\right)$ and $\tilde{\pi}_{m}\left(v\right)$,
compute the soft frequencies of ratings for each user/movie group
pair as

\[
w(r|u,v)=\frac{{\displaystyle \sum_{(n,m)\in O:R_{nm}=r}}\pi_{n}\left(u\right)\tilde{\pi}_{m}\left(v\right)}{{\displaystyle \sum_{r\in\mathcal{R}}}{\displaystyle \sum_{(n,m)\in O:R_{nm}=r}}\pi_{n}\left(u\right)\tilde{\pi}_{m}\left(v\right)}.\]

\end{algorithm*}
The exact update equations are given in Algorithm \ref{alg:IMP}.
Though the idea is similar to an EM update, the resulting equation
are different and seem to perform much better.

\subsection{Expectation Maximization (EM) Learning }

\noindent Now, we reformulate the problem in a standard variational
EM framework and propose a second algorithm by minimizing an upper
bound on the free energy \cite{Neal Graph}. In other words, we view
the problem as maximum-likelihood parameter estimation problem where
$p_{U_{n}}(\cdot)$, $p_{V_{m}}(\cdot)$, and $p_{R|U,M}(\cdot|\cdot)$
are the model parameters $\theta$ and $\mathbf{U},\mathbf{V}$ are
the missing data. For each of these parameters, the $i$-th estimate
is denoted $f_{n}^{(i)}(u)$, $h_{m}^{(i)}(v)$, and $w^{(i)}(r|u,v)$.
Let $O\subseteq[N]\times[M]$ be the set of user-movie pairs that
have been observed. Then, we can write the complete data (negative)
log-likelihood as \begin{align*}
R^{c}\left(\theta\right) & =-\mbox{log}\!\!\!\prod_{(n,m)\in O}\!\!\!\Pr\left(R_{nm}=r_{n,m},U_{n}=u_{n},V_{m}=v_{m}\right)\\
 & =-\mbox{log}\,\prod_{(n,m)\in O}w\left(r_{n,m}|u_{n},v_{m}\right)f_{n}\left(u_{n}\right)h_{m}\left(v_{m}\right).\end{align*}
Using a variational approach, this can be upper bounded by\[
\sum_{(n,m)\in O}\!\!\!\!\!\! D\left(Q_{U_{n},V_{n}|R_{nm}}(\cdot,\cdot|r_{n,m})||\hat{p}_{U_{n},V_{m}|R_{nm}}(\cdot,\cdot|r_{n,m})\right),\]

\noindent where we introduce the variational probability distributions
$Q_{U_{n},V_{m}|R_{nm}}\left(u,v|r\right)$ that satisfy \[
\sum_{u,v}Q_{U_{n},V_{m}|R_{nm}}\left(u,v|r\right)=1\]
 and let \[
\hat{p}_{U_{n},V_{m}|R_{nm}}(u,v|r)=\frac{w\left(r_{n,m}|u,v\right)f_{n}\left(u\right)h_{m}\left(v\right)}{\sum_{u',v'}w\left(r_{n,m}|u',v'\right)f_{n}\left(u'\right)h_{m}\left(v'\right)}.\]

\noindent The variational EM algorithm we have developed uses alternating
steps of KL divergence minimization to estimate the underlying generative
model \cite{Csiszar alter}. The results show that this variational
approach gives the equivalent update rule as the standard EM framework
(with a simpler derivation in Appendix \ref{sec:AppendixEM}) which
guarantees convergence to local minima. The update equations are presented
in Algorithm \ref{alg:EM_alg}. This learning algorithm, in fact,
extends Thomas Hofmann's work and generalizes probabilistic matrix
factorization (PMF) results \cite{Hofmann uai99,PMF2}. Its main drawback
is that it is difficult to analyze because the effects of initial
conditions and local minima can be very complicated.

\subsection{\label{sec:Prediction_sec}Prediction and Initialization }

\noindent Since the primary goal is the prediction of hidden variables
based on observed ratings, the learning algorithms focus on estimating
the distribution of each hidden variable given the observed ratings.
In particular, the outputs of both algorithms (after $i$ iterations)
are estimates of the distributions for $R_{nm}$, $U_{n}$, and $V_{m}$.
They are denoted, respectively, ${\textstyle \hat{p}_{R_{nm}|\mathbf{R}_{O}}^{(i+1)}(r)}$,
${\textstyle \hat{p}_{U_{n}|\mathbf{R}_{O}}^{(i+1)}(u)}$, and $\hat{p}_{V_{m}|\mathbf{R}_{O}}^{(i+1)}(v)$.
Using these, one can minimize various types of prediction error. For
example, minimizing the mean-squared prediction error results in the
conditional mean estimate\[
\hat{r}_{n,m,1}^{(i)}=\sum_{r\in\mathcal{R}}r\,{\textstyle \hat{p}_{R_{nm}|\mathbf{R}_{O}}^{(i)}(r)}.\]
While minimizing the classification error of users (and movies) into
groups results in the maximum a posteriori (MAP) estimates\begin{eqnarray*}
\hat{u}_{n}^{(i)}=\arg\max_{u}{\textstyle \hat{p}_{U_{n}|\mathbf{R}_{O}}^{(i)}(u)} &  & \hat{v}_{m}^{(i)}=\arg\max_{v}{\textstyle \hat{p}_{V_{m}|\mathbf{R}_{O}}^{(i)}(v)}.\end{eqnarray*}
Likewise, after $w^{(i)}\left(r|u,v\right)$ converges, we can also
make hard decisions on groups by the MAP estimates first and then
compute rating predictions by 

\[
\hat{r}_{n,m,2}^{(i)}=\sum_{r\in\mathcal{R}}r\, w^{(i)}\left(r|\hat{u}_{n}^{(i)},\hat{v}_{m}^{(i)}\right).\]
While this should perform worse with exact inference, this may not
be the case with approximate inference algorithms.

Both iterative learning algorithms require proper initial estimates
of a set of initial group (of the user and movie) probabilities $f_{n}^{(0)}\left(u\right),\, h_{m}^{(0)}(v)$
and observation model$,\, w^{(0)}(r|u,v)$ since randomized initialization
often leads to local minima and poor performance. To cluster users
(or movies), we employ a variable-dimension vector quantization (VDVQ)
algorithm \cite{Gersho} and the standard codebook splitting approach
known as the generalized Lloyd algorithm (GLA) to generate codebooks
whose size is any power of 2. The VDVQ algorithm is essentially based
on alternating minimization of the \textcolor{black}{average distance
between users (or movies) and codebooks} (that contains no missing
data) with the two optimality criteria: \emph{nearest neighbor} and
\emph{centroid} rules only on the elements both vectors share. The
group probabilities are initialized by assuming that the VDVQ gives
the {}``correct'' group with probability $\epsilon=0.9$ and spreads
its errors uniformly across all other groups. In the case of users,
one can think of this Algorithm \ref{alg:CR_alg} as a {}``$k$-critics''
algorithms which tries to design $k$ critics (i.e., people who have
seen every movie) that cover the space of all user tastes and each
user is given a soft {}``degree of assignment (or soft group membership)''
to each of the critics which can take on values between 0 and 1.

\section{Density Evolution Analysis \label{sec:DE_sec}}

\noindent Density evolution (DE) is well-known technique for analyzing
probabilistic message-passing inference algorithms that was originally
developed to analyze belief-propagation decoding of error-correcting
codes and was later extended to more general inference problems \cite{SparseObs}.
It works by tracking the distribution of the messages passed in the
graph under the assumption that the local neighborhood of each node
is a tree. While this assumption is not rigorous, it is motivated
by the following lemma. We consider the factor graph for a randomly
chosen instance of this problem. The key assumption is that the outgoing
edges from each user node are attached to movie nodes via a random
permutation. This is identical to the model used for irregular LDPC
codes \cite{RU2001}. 
\begin{lemma}
\label{lem:tree}Let $\mathcal{N}_{l}(v)$ denote the depth-$l$ neighborhood
(i.e., the induced subgraph including all nodes within $l$ steps
from $v$) of an arbitrary user (or movie) node $v$. Let the problem
size $N$ become unbounded with $M=\beta N$ for $\beta<1$, maximum
degree $d_{N}$, and depth-$l_{N}$ neighborhoods. One finds that
if \[
\frac{(2l_{N}+1)\log d_{N}}{\log N}<1-\delta,\]
for some $\delta>0$ and all $N$, then the graph $\mathcal{N}_{l}(v)$
is a tree w.h.p. for almost all $v$ as $N\rightarrow\infty$.\end{lemma}
\begin{proof}
The proof follows from a careful treatment of standard tree-like neighborhood
arguments as in Appendix \ref{sec:Treelemma}.
\end{proof}
For this problem, the messages passed during inference consist of
belief functions for user groups (e.g., passed from movie nodes to
user nodes) and movie groups (e.g., passed form user nodes to movie
nodes). The message set for user belief functions is $\mathcal{M}_{u}=\mathcal{P}([g_{u}])$,
where $\mathcal{P}(S)$ is the set of probability distributions over
the finite set $S$. Likewise, the message set for movie belief functions
is $\mathcal{\mathcal{M}}_{v}=\mathcal{P}([g_{v}])$. The decoder
combines $d$ user (resp. movie) belief-functions $a_{1}(\cdot),\ldots,a_{d}(\cdot)\in\mathcal{M}_{u}$
(resp. $b_{1}(\cdot),\ldots,b_{d}(\cdot)\in\mathcal{M}_{v}$) using
\begin{align*}
F_{d}\left(a_{1},r_{1},...,a_{d},r_{d};b\right)\! & \triangleq\!\frac{b(v)\prod_{j=1}^{d}\sum_{u}\! a_{j}(u)w\left(r_{j}|u,v\right)}{\sum_{v}b(v)\prod_{j}\sum_{u}\! a_{j}(u)w\left(r_{j}|u,v\right)}\\
G_{d}\left(b_{1},r_{1},...,b_{d},r_{d};a\right)\! & \triangleq\!\frac{a(u)\prod_{j=1}^{d}\sum_{v}\! b_{j}(v)w\left(r_{j}|u,v\right)}{\sum_{u}a(u)\prod_{j}\sum_{v}\! b_{j}(v)w\left(r_{j}|u,v\right)}.\end{align*}

Since we need to consider the possibility that the ratings are generated
by a process other than the assumed model, we must also keep track
of the true user (or movie) group associated with each belief function.
Let $\mu^{(i)}(u,A)$ (resp. $\nu^{(i)}(v,B)$) be the probability
that, during the $i$-th iteration, a randomly chosen user (resp.
movie) message is coming from a node with true user group $u$ (resp.
movie group $v$) and has a user belief function $a(\cdot)\in A\subseteq\mathcal{M}_{u}$
(resp. movie belief function $b(\cdot)\in B\subseteq\mathcal{M}_{v}$).
The DE update equations for degree $d$ user and movie nodes, in the
spirit of \cite{SparseObs}, are shown in equations (\ref{eq:User_DE})
and (\ref{eq:Movie_DE}) where $I(x\in A)$ is defined as a indicator
function \[
I(x\in A)=\begin{cases}
1 & \mbox{if}\, x\in A\\
0 & \mbox{if}\, x\notin A\end{cases}.\]

\begin{figure*}
\lyxline{\normalsize}

\begin{align}
\mu_{d}^{(i+1)}\!(u,B) & \!=\!\int\sum_{r_{1},\ldots,r_{d}}I\left(G\left((b_{1},r_{1}),\ldots,(b_{d},r_{d});a\right)\!\in\! B\right)\mu^{(0)}\!\left(u,da\right)\prod_{j=1}^{d}\sum_{v}\nu^{(i)}\!\left(v,db_{j}\right)w\left(r_{j}|u,v\right)\label{eq:User_DE}\\
\nu_{d}^{(i+1)}\!(v,A) & \!=\!\int\sum_{r_{1},\ldots,r_{d}}I\left(F\left((a_{1},r_{1}),\ldots,(a_{d},r_{d});b\right)\!\in\! A\right)\nu^{(0)}\!\left(v,db\right)\prod_{j=1}^{d}\sum_{u}\mu^{(i)}\!\left(u,da_{j}\right)w\left(r_{j}|u,v\right)\label{eq:Movie_DE}\end{align}

\lyxline{\normalsize}
\end{figure*}

Like LDPC codes, we expect to see that the performance of Algorithm
\ref{alg:IMP} depends crucially on the degree structure of the factor
graph. Therefore, we let $\Lambda_{j}$ (resp. $\Gamma_{j}$) be the
fraction of user (resp. movie) nodes with degree $j$ and define the
edge degree distribution to be $\lambda_{j}=\Lambda_{j}j/\sum_{k\geq1}\Lambda_{k}k$
(resp.$\rho_{j}=\Gamma_{j}j/\sum_{k\geq1}\Gamma_{k}k$). Averaging
over the degree distribution gives the final update equations

\begin{align*}
\mu^{(i+1)}(u,B) & =\sum_{d\geq1}\lambda_{d}\mu_{d}^{(i+1)}(u,B)\\
\nu^{(i+1)}(v,A) & =\sum_{d\geq1}\rho_{d}\nu_{d}^{(i+1)}(v,A).\end{align*}
We anticipate that this analysis will help us understand the IMP algorithm's
observed performance for large problems based on the success of DE
for channel coding problems.

\section{Experimental Results \label{sec:Simulations}}

\begin{figure*}[p]
\begin{centering}
\includegraphics[width=0.7\columnwidth]{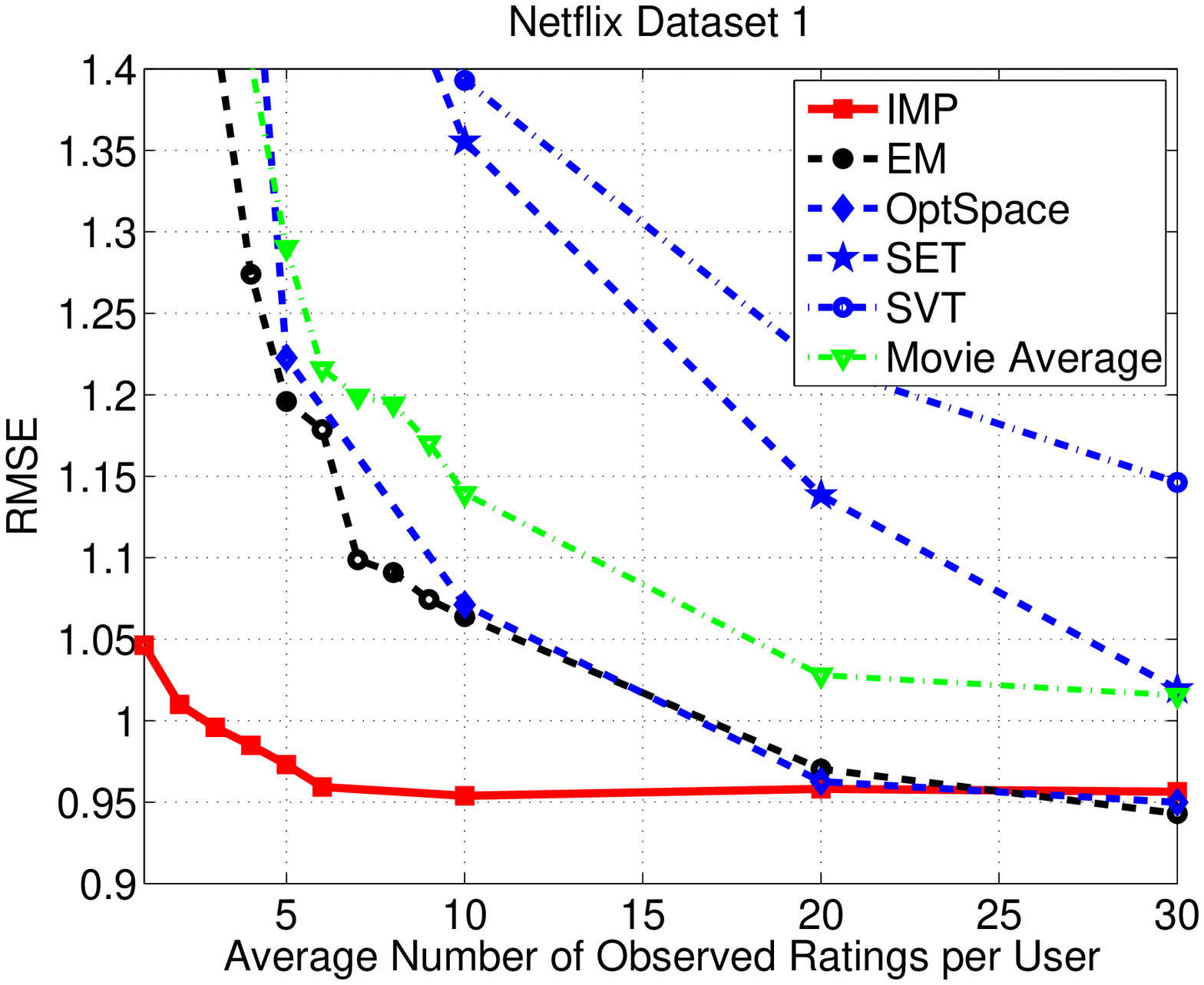}
\par\end{centering}

\begin{centering}
\includegraphics[width=0.7\columnwidth]{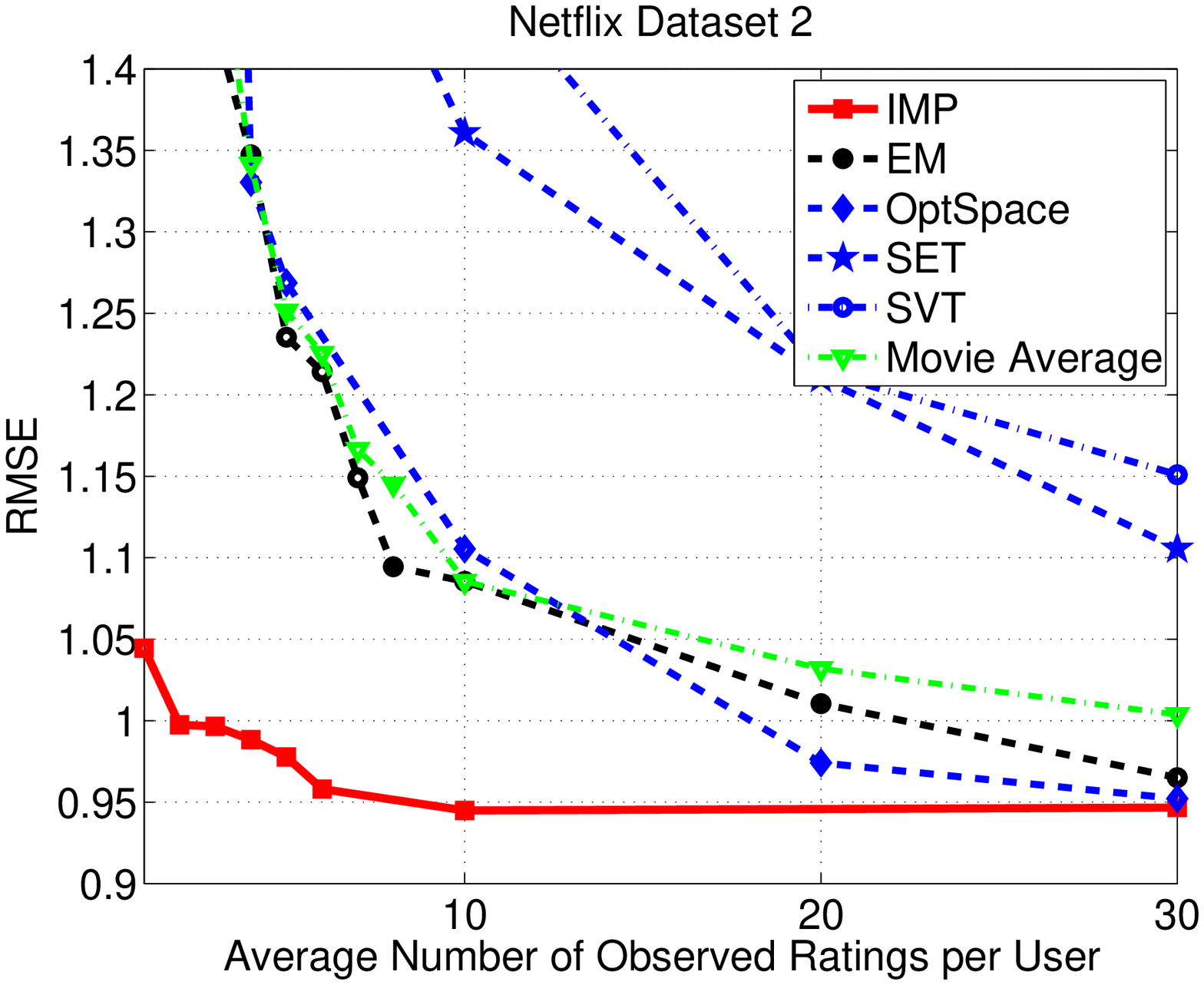}
\par\end{centering}

\caption{\label{fig:netflix}Remedy for the Cold-Start Problem: Each plot shows
the RMSE on the validation set versus the average number of observations
per user for Netflix datasets. Performance is compared with three
different matrix completion algorithms (OptSpace \cite{OptSpace},
SET \cite{SET} and SVT \cite{SVT}) and an algorithm that uses the
average rating for each movie as the prediction. For IMP and EM, $\hat{r}_{n,m,1}^{(i)}$
prediction formula is used.}

\end{figure*}
\begin{figure*}
\begin{centering}
\includegraphics[width=0.7\columnwidth]{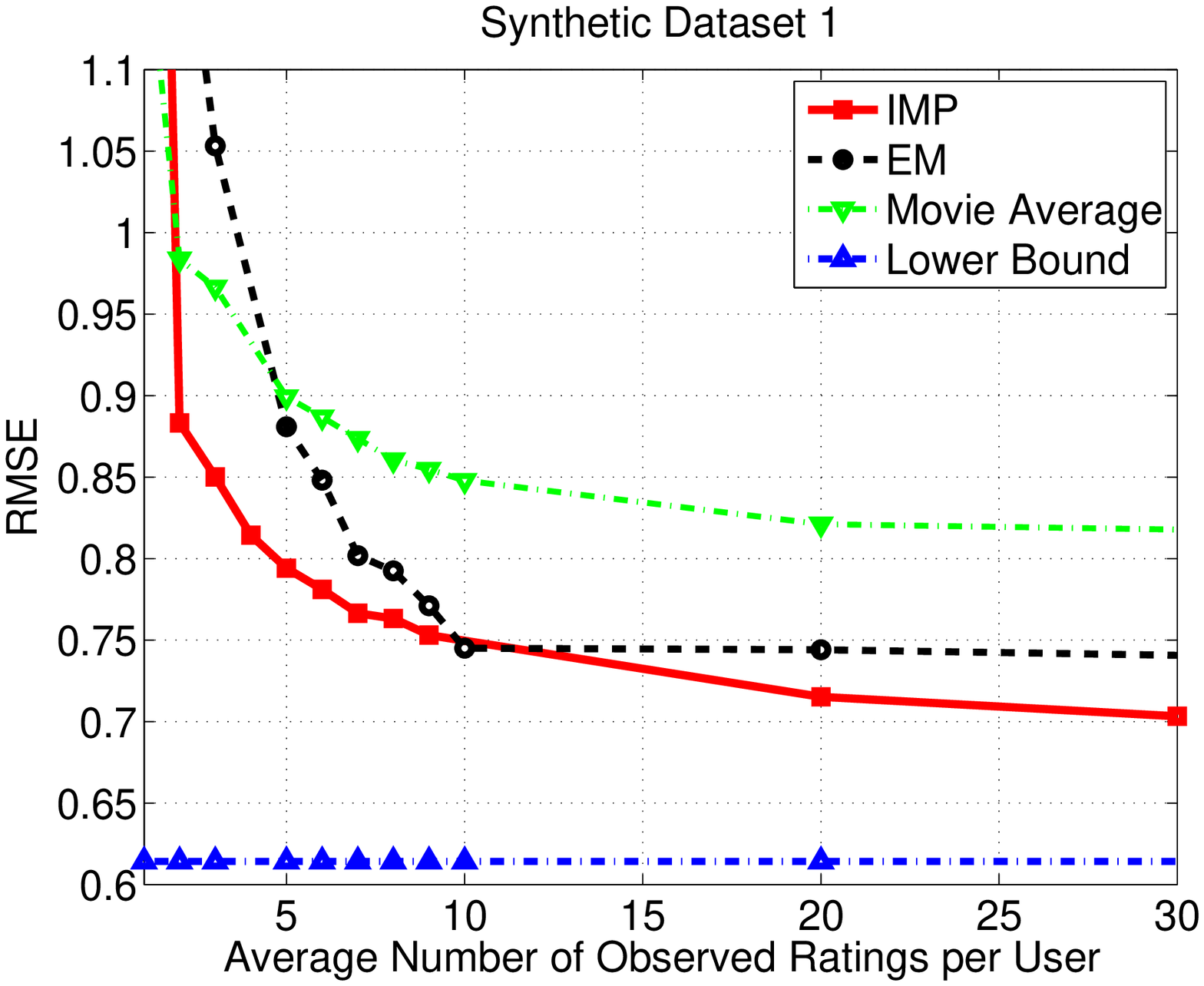}
\par\end{centering}

\centering{}\includegraphics[width=0.7\columnwidth]{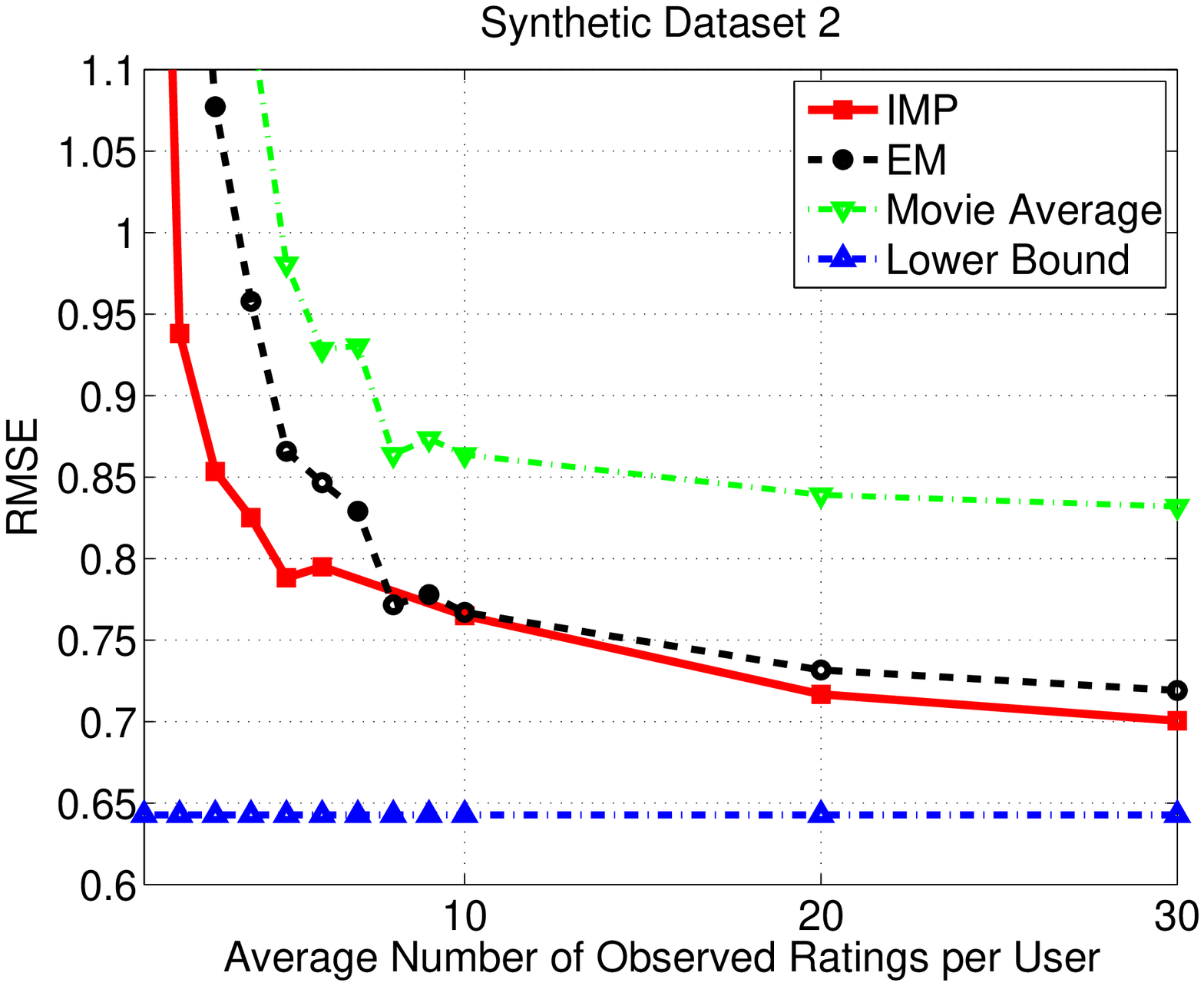}\caption{\label{fig:synthetic}Each plot shows the RMSE on the validation set
versus the average number of observations per user for synthetic datasets.
Performance is compared with an (analytical) lower bound on RMSE assuming
known user and movie group. }

\end{figure*}

\subsection{Details of Datasets and Training}

\noindent The key challenge of collaborative filtering problem is
predicting the preference of a user for a given item based only on
very few known ratings in a way that minimizes some per-letter metric
$d(r,r')$ for ratings. To study this, we created two smaller datasets
from the Netflix data by randomly subsampling user/movie/rating triples
from the original Netflix datasets which emphasizes the advantages
of MP scheme. This idea was followed from \cite{PMF2,LowRank}.
\begin{itemize}
\item \vspace{1mm}\textbf{Netflix Dataset 1} is a matrix given by the first
5,000 movies and users. This matrix contains 280,714 user/movie pairs.
Over 15\% of the users and 43\% of the movies have less than 3 ratings. 
\item \textbf{Netflix Dataset 2} is a matrix of 5,035 movies and 5,017 users
by selecting some 5,300 movies and 7,250 users and avoiding movies
and users with less than 3 ratings. This matrix contains 454,218 user/movie
pairs. Over 16\% of the users and 41\% of the movies have less than
10 ratings. \vspace{1mm}
\end{itemize}
\noindent To provide further insights into the quality of the proposed
factor graph model and suboptimality of the algorithms by comparison
with the theoretical lower bounds, we generated two synthetic datasets
from the above partial matrices. The synthetic datasets are generated
once with the learned density ${\textstyle \hat{p}_{R_{nm}|\mathbf{R}_{O}}^{(i)}(r)}$,
${\textstyle \hat{p}_{U_{n}|\mathbf{R}_{O}}^{(i)}(u)}$, and $\hat{p}_{V_{m}|\mathbf{R}_{O}}^{(i)}(v)$
and then randomly subsampled as the partial Netflix datasets.
\begin{itemize}
\item \vspace{1mm}\textbf{Synthetic Dataset 1} is generated after learning
Netflix Dataset 1 with $g_{u}\!=\! g_{v}\!=\!8$.
\item \textbf{Synthetic Dataset 2} is generated after learning Netflix Dataset
2 with $g_{u}\!=\! g_{v}\!=\!16$.\vspace{1mm}
\end{itemize}
Additionally, to evaluate the performance of different algorithms/models
efficiently, we hide 1,000 randomly selected user/movie entries as
a validation set from each dataset. Note that the choice of $g_{u}$
and $g_{v}$ to obtain synthetic datasets resulted in the competitive
performance on this validation set, but not fully optimized. Simulations
were performed on these partial datasets where the average number
of observed ratings per user was varied between 1 and 30. The experimental
results are shown in Fig. \ref{fig:netflix} and \ref{fig:synthetic}
and the performance is evaluated using the root mean squared error
(RMSE) of prediction defined by \[
\sqrt{\sum_{(n,m)\in S}\left(\hat{r}_{n,m}-r_{n,m}\right)^{2}/\left|S\right|}.\]

\subsection{Results and Model Comparisons\label{subsec:Discusssions}}

While the IMP algorithm is not yet competitive on the entire Netflix
dataset \cite{netflix}, however, it shows some promise for the recommender
systems based on MP frameworks. In reality, we have discovered that
MP approaches really do improve the cold-start problem. Here is a
summary of observations we've learned from the simulation study.
\begin{enumerate}
\item \textbf{Improvement of the cold-start problem with MP algorithms}:
From Fig. \ref{fig:netflix} results on partial Netflix datasets,
we clearly see while many methods perform similarly with large amounts
of observed ratings, IMP is superior for small amounts of data. This
better threshold performance of the IMP algorithm over the other algorithms
does help reduce the cold start problem. This provides strong support
to use MP approaches in standard CF systems. Also in simulations,
we observe lower computational complexity of the IMP algorithm even
though we have developed computationally efficient versions of our
EM algorithm (see Appendix \ref{sec:AppendixEM}). 
\item \textbf{Comparison with low-rank matrix models}: Our factor graph
model is a probabilistic generalization of other low-rank matrix models.
Similar asymptotic behavior (for enough measurements) between partial
Netflix and synthetic dataset suggests that the factor graph model
is a good fit for this problem. By comparing with the results in \cite{LowRank},
we can support that the Netflix dataset is much well described by
the factor graph model. Other than these advantages, each output group
has generative nature with explicit semantics. In other words, after
learning the density, we can use them to generate synthetic data with
clear meanings. These benefits do not extend to low-rank matrix models
easily.
\end{enumerate}

\section{Conclusions}

\noindent For the Netflix problem, a number of researchers have used
low-rank models that lead to learning methods based on SVD and principal
component analysis (PCA) with a least squares flavor. Unlike these
prior works, in this paper, we proposed the new factor graph model
and successfully applied MP framework to the problem. First, we presented
the IMP algorithm and used simulations to show its superiority over
other algorithms by focusing on the cold-start problem. Second, we
studied quality of the model by deriving the DE analysis with a generalization
error bound and complementing these theoretical analyses with simulation
results for synthetic data generated from the learned model. 

\pagebreak{}\appendices

\section{Proof of Theorem \ref{thm:bound}\label{sec:Bound}}

This proof follows arguments of the generalization error in \cite{Nati}.
First, fix $Y$ as well as $X\in R^{N\times M}$. When an index pair
$\left(n,\, m\right)$ is chosen uniformly random, $\mbox{d}\left(x_{n,m},\, y_{n,m}\right)$
is a Bernoulli random variable with probability $D\left(X,\, Y\right)$
of being one. If the entries of $O$ are chosen independently random,
$|O|D_{O}\left(X,\, Y\right)$ is binomially distributed with parameters
$|O|D\left(X,\, Y\right)$ and $|O|\epsilon$. Using Chernoff's inequality,
we get \begin{align*}
\Pr\left(D\left(X,\, Y\right)\geq D_{O}\left(X,\, Y\right)+\epsilon\right) & =\Pr\left(|O|D_{O}\left(X,\, Y\right)\leq|O|D\left(X,\, Y\right)-|O|\epsilon\right)\leq e^{-2|O|\epsilon^{2}}.\end{align*}
Now note that $\mbox{d}\left(x,\, y\right)$ only depends on the sign
of $xy$, so it is enough to consider equivalence classes of matrices
with the same sign patterns. Let $f\left(N,\, M,\, g_{u},\, g_{v}\right)$
be the number of such equivalence classes. For all matrices in an
equivalence class, the random variable $D_{O}\left(X,\, Y\right)$
is the same. Thus we take a union bound of the events $\left\{ X|D\left(X,\, Y\right)\geq D_{O}\left(X,\, Y\right)+\epsilon\right\} $
for each of these $f\left(N,\, M,\, g_{u},\, g_{v}\right)$ random
variables with the bound above and $\epsilon=\sqrt{\frac{\mbox{log}\, f\left(N,\, M,\, g_{u},\, g_{v}\right)-\mbox{log}_{}\delta}{2|O|}}$,
we have\[
\Pr\left(\exists X\in\chi_{g_{u},g_{v}}\, D\left(X,\, Y\right)\geq D_{O}\left(X,\, Y\right)+\sqrt{\frac{\mbox{log}\, f\left(N,\, M,\, g_{u},\, g_{v}\right)-\mbox{log}\delta}{2|O|}}\right)\leq\delta.\]
Since any matrix $X\in\chi_{g_{u},g_{v}}$ can be written as $X=U^{T}GV$,
to bound the number of sign patterns of $X$,~$f\left(N,\, M,\, g_{u},\, g_{v}\right)$,
consider $Ng_{u}+Mg_{v}+g_{u}g_{v}$ entries of $U,\, G,\, V$ as
variables and the $NM$ entries of $X$ as polynomials of degree three
over these variables as\[
x_{n,m}=\sum_{k=1}^{g_{u}}\sum_{l=1}^{g_{v}}u_{k,n}\cdot g_{k,l}\cdot v_{l,m}.\]
By the use of the bound in lemma 2, we obtain \[
f\left(N,\, M,\, g_{u},\, g_{v}\right)\leq\left(\frac{4e\cdot3\cdot NM}{Ng_{u}+Mg_{v}+g_{u}g_{v}}\right)^{Ng_{u}+Mg_{v}+g_{u}g_{v}}\leq\left(\frac{12eM}{\mbox{min}(g_{u},\, g_{v})}\right)^{Ng_{u}+Mg_{v}+g_{u}g_{v}}.\]
This bound yields a factor of $\mbox{log}\,\frac{12eM}{\mbox{min}(g_{u},\, g_{v})}$
in the bound and establishes the theorem. 
\begin{lemma}
\cite{Alon} Total number of sign patterns of $r$ polynomials, each
of degree at most $d$, over $q$ variables, is at most $\left(8edr/q\right)^{q}$
if $2r>q>2$. Also, total number of sign patterns of $r$ polynomials
with $\left\{ \pm1\right\} $ coordinates, each of degree at most
$d$, over $q$ variables, is at most $\left(4edr/q\right)^{q}$ if
$r>q>2$. 
\end{lemma}

\section{Derivation of Algorithm \ref{alg:EM_alg} \label{sec:AppendixEM}}

As the first step, we specify a complete data likelihood as

\[
\Pr\left(R_{nm}=r_{n,m},U_{n}=u_{n},V_{m}=v_{m}\right)=w\left(r_{n,m}|u_{n},v_{m}\right)f_{n}\left(u_{n}\right)h_{m}\left(v_{m}\right)\]
and the corresponding (negative) log-likelihood function can be written
as\begin{align*}
R^{c}\left(\theta\right) & =-\mbox{log}\!\!\!\prod_{(n,m)\in O}\!\!\!\Pr\left(R_{nm}=r_{n,m},U_{n}=u_{n},V_{m}=v_{m}\right)\\
 & =-\sum_{(n,m)\in O}\left[\mbox{log}\, w\left(r_{n,m}|u_{n},v_{m}\right)+\mbox{log}\, f_{n}\left(u_{n}\right)+\mbox{log}\, h_{m}\left(v_{m}\right)\right]\end{align*}
The variational EM algorithm now consists of two steps that are performed
in alternation with a Q distribution to approximate a general distribution.

\subsection{E-step}

Since the states of the latent variables are not known, we introduce
a variational probability distribution\[
Q_{U_{n},V_{m}|R_{nm}}\left(u,v|r\right)\,\mbox{subject\,\ to}\,\sum_{u,v}Q_{U_{n},V_{m}|R_{nm}}\left(u,v|r\right)=1\]
for all observed pairs $(n,m)$. Exploiting the concavity of the logarithm
and using Jensen's inequality, we have\begin{align*}
R\left(\theta\right) & =-\sum_{(n,m)\in O}\mbox{log\,}\sum_{u,v}\,\Pr\left(R_{nm}=r_{n,m},U_{n}=u_{n},V_{m}=v_{m}\right)\\
 & =-\sum_{(n,m)\in O}\mbox{log\,}\sum_{u,v}\, Q_{U_{n},V_{m}|R_{nm}}\left(u,v|r\right)\frac{w\left(r_{n,m}|u,v\right)f_{n}\left(u\right)h_{m}\left(v\right)}{Q_{U_{n},V_{m}|R_{nm}}\left(u,v|r\right)}\\
 & \leq-\sum_{(n,m)\in O}\sum_{u,v}\, Q_{U_{n},V_{m}|R_{nm}}\left(u,v|r\right)\mbox{log\,}\frac{w\left(r_{n,m}|u,v\right)f_{n}\left(u\right)h_{m}\left(v\right)}{Q_{U_{n},V_{m}|R_{nm}}\left(u,v|r\right)}\\
 & \triangleq\bar{R}\left(\theta\left|Q\right.\right)-\sum_{(n,m)\in O}H\left(Q\left(\cdot|u,v,r\right)\right)\\
 & \triangleq R\left(\theta;\, Q\right)\end{align*}
To compute the tightest bound given parameters $\hat{\theta}$ i.e.,
we optimize the bound w.r.t the $Q$s using\[
\nabla_{Q}\left[R\left(\theta;\, Q\right)+\sum_{(n,m)\in O}\sum_{u,v}\lambda_{u,v}\, Q\right]=0.\]
These yield posterior probabilities of the latent variables, \[
\hat{p}_{U_{n},V_{m}|R_{nm}}(u,v|r;\hat{\theta})=Q_{U_{n},V_{m}|R_{nm}}^{*}\left(u,v|r;\hat{\theta}\right)=\frac{w\left(r_{n,m}|u,v\right)f_{n}\left(u\right)h_{m}\left(v\right)}{\sum_{u',v'}w\left(r_{n,m}|u',v'\right)f_{n}\left(u'\right)h_{m}\left(v'\right)}.\]
Also note that we can get the same result by Gibbs inequality as\begin{align*}
R\left(\theta\right) & \leq-\sum_{(n,m)\in O}\sum_{u,v}\, Q_{U_{n},V_{m}|R_{nm}}\left(u,v|r\right)\mbox{log\,}\frac{w\left(r_{n,m}|u,v\right)f_{n}\left(u\right)h_{m}\left(v\right)}{Q_{U_{n},V_{m}|R_{nm}}\left(u,v|r\right)}\\
 & =\sum_{(n,m)\in O}\!\!\!\!\!\! D\left(Q_{U_{n},V_{n}|R_{nm}}(\cdot,\cdot|r_{n,m})||\hat{p}_{U_{n},V_{m}|R_{nm}}(\cdot,\cdot|r_{n,m})\right).\end{align*}

\subsection{M-step}

Obviously the posterior probabilities need only to be computed for
pairs $\left(n,\, m\right)$ that have actually been observed. Thus
optimize \begin{align*}
\bar{R}\left(\theta,\hat{\theta}\right) & =-\sum_{(n,m)\in O}\sum_{u,v}\, Q_{U_{n},V_{m}|R_{nm}}^{*}\left(u,v|r;\hat{\theta}\right)\left\{ \mbox{log}\, w\left(r_{n,m}|u,v\right)+\mbox{log}\, f_{n}\left(u\right)+\mbox{log}\, h_{m}\left(v\right)\right\} \\
 & =-\sum_{(n,m)\in O}\sum_{u,v}\,\frac{w\left(r_{n,m}|u,v\right)f_{n}\left(u\right)h_{m}\left(v\right)}{\sum_{u',v'}w\left(r_{n,m}|u',v'\right)f_{n}\left(u'\right)h_{m}\left(v'\right)}\left\{ \mbox{log}\, w\left(r_{n,m}|u,v\right)+\mbox{log}\, f_{n}\left(u\right)+\mbox{log}\, h_{m}\left(v\right)\right\} \end{align*}
with respect to parameters $\theta$ which leads to the three sets
of equations for the update of \[
w\left(r|u,v\right),\, f_{n}\left(u\right),\, h_{m}\left(v\right).\]
Moreover, for large scale problems, to avoid computational loads of
each step, combining both E and M steps by plugging $Q$ function
into M-step gives more tractable EM Algorithm. The resulting equations
are defined in Algorithm \ref{alg:EM_alg}.

\section{Proof of Lemma \ref{lem:tree}\label{sec:Treelemma}}

Starting from any node $v$, we can recursively grow $\mathcal{N}_{i+1}(v)$
from $\mathcal{N}_{i}(v)$ by adding all neighbors at distance $i+1$.
Let $A_{i}$ be the number of outgoing edges from $\mathcal{N}_{i}(v)$
to the next level and $b_{1}^{(i)},\ldots,b_{n}^{(i)}$ be the degrees
of the $n_{i}$ available nodes that can be chosen in the next level.
The probability that the graph remains a tree is \[
p\left(A_{i},\mathbf{b}^{(i)}\right)=\frac{\sum_{S\subset[n],|S|=A_{i}}\prod_{s\in S}b_{s}^{(i)}}{\binom{\sum_{j=1}^{n}b_{j}^{(i)}}{A_{i}}},\]
where the numerator is the number of ways that the $A_{i}$ edges
can attach to distinct nodes in the next level and the denominator
is the total number of ways that the $A_{i}$ edges may attach to
the available nodes. Using the fact that the numerator is an unnormalized
expected value of the product of $A_{i}$ $b$'s drawn without replacement,
we can lower bound the numerator using \[
\sum_{S\subset[n],|S|=A_{i}}\prod_{s\in S}b_{s}^{(i)}\geq\binom{n_{i}}{A_{i}}\left(\overline{b}_{i}-\frac{(d-1)A_{i}}{n_{i}}\right)^{A_{i}}\geq\frac{\left(n_{i}-A_{i}\right)^{A_{i}}}{A_{i}!}\left(\overline{b}_{i}-\frac{(d-1)A_{i}}{n_{i}}\right)^{A_{i}}.\]
This can be seen as lower bounding the expected value of $A_{i}$
$b$'s drawn from with replacement from a distribution with a slightly
lower mean. Upper bounding the denominator by $(n_{i}\overline{b}_{i})^{A_{i}}/A_{i}!$
gives \begin{align*}
p\left(A_{i},\mathbf{b}^{(i)}\right) & \geq\frac{\left(n_{i}-A_{i}\right)^{A_{i}}A_{i}!\left(\overline{b}_{i}-\frac{(d-1)A_{i}}{n_{i}}\right)^{A_{i}}}{\left(n_{i}\overline{b}_{i}\right)^{A_{i}}A_{i}!}\\
 & =\left(1-\frac{A_{i}}{n_{i}}\right)^{A_{i}}\left(1-\frac{(d-1)A_{i}}{\overline{b}_{i}n_{i}}\right)^{A_{i}}\\
 & \geq\left(1-\frac{A_{i}^{2}}{n_{i}}-\frac{A_{i}^{2}(d-1)}{\overline{b}_{i}n_{i}}\right).\end{align*}
Now, we can take the product from $i=0,\ldots,l-1$ to get\begin{align*}
\Pr\left(\mathcal{N}_{l}(v)\mbox{ is a tree}\right) & =\prod_{i=0}^{l-1}\Pr\left(\mathcal{N}_{i+1}(v)\mbox{ is a tree}|\mathcal{N}_{0}(v),\ldots,\mathcal{N}_{i}(v)\mbox{ are trees}\right)\\
 & \geq\prod_{i=0}^{l-1}\left(1-\frac{A_{i}^{2}}{n_{i}}-\frac{A_{i}^{2}(d-1)}{\overline{b}_{i}n_{i}}\right)\\
 & \geq1-\sum_{i=0}^{l-1}\left(\frac{A_{i}^{2}}{n_{i}}+\frac{A_{i}^{2}(d-1)}{\overline{b}_{i}n_{i}}\right)\\
 & \geq1-\left(1+\frac{1}{d^{2}-1}\right)\left(\frac{d^{2l}}{\beta N-d^{l}}+\frac{d^{2l}(d-1)}{\beta N-d^{l}}\right)\\
 & \geq1-\left(1+\frac{1}{d^{2}-1}\right)\frac{d^{2l+1}}{\beta N-d^{l}},\end{align*}
because $A_{i}\leq d^{i+1}$, $\sum_{i=0}^{l-1}A_{i}^{2}\leq d^{2}\frac{d^{2l}}{d^{2}-1}=d^{2l}\left(1+\frac{1}{d^{2}-1}\right)$,
and $n_{i}\geq\beta N-\sum_{j=0}^{i}d^{j}\geq\beta N-d^{i+1}$. Examining
the expression \[
\log\frac{d^{2l+1}}{\beta N-d^{l}}\leq(2l_{N}+1)\log d_{N}-\log N+O(1)\leq-\delta\log N+O(1)\]
shows that the probability of failure is $O\left(N^{-\delta}\right)$. 

Let $Z$ be a r.v. whose value is the number of user nodes whose depth-$l$
neighborhood is not a tree. We can upper bound the expected value
of $Z$ with\[
E[Z]\leq\frac{d^{2l+1}}{\Theta(N)-d^{l}}N\leq\frac{O\left(N^{-\delta}\right)}{\Theta\left(N\right)-O\left(N^{1/2}\right)}N=O\left(N^{1-\delta}\right).\]
With Markov's inequality, one can show that \[
\Pr\left(Z\geq N^{1-\delta/2}\right)\leq\frac{E[Z]}{N^{1-\delta/2}}\leq\frac{O\left(N^{1-\delta}\right)}{N^{1-\delta/2}}.\]
Therefore, the depth-$l$ neighborhood is a tree (w.h.p. as $N\rightarrow\infty$)
for all but a vanishing fraction of user nodes.
\end{document}